# TREES: A CPU/GPU Task-Parallel Runtime with Explicit Epoch Synchronization

Blake A. Hechtman, Andrew D. Hilton, and Daniel J. Sorin
Department of Electrical and Computer Engineering
Duke University

*Abstract*—We have developed a task-parallel runtime system, called TREES, that is designed for high performance on CPU/GPU platforms. On platforms with multiple CPUs, Cilk's "work-first" principle underlies how task-parallel applications can achieve performance, but work-first is a poor fit for GPUs. We build upon work-first to create the "work-together" principle that addresses the specific strengths and weaknesses of GPUs. The work-together principle extends work-first by stating that (a) the overhead on the critical path should be paid by the entire system at once and (b) work overheads should be paid co-operatively. We have implemented the TREES runtime in OpenCL, and we experimentally evaluate TREES applications on a CPU/GPU platform.

## 1. Introduction

GPUs support data parallelism efficiently, but not all problems are data-parallel. Non-data-parallel problems, which are also called irregular parallel problems, often stress systems to balance load, resolve dependencies, and enforce memory consistency. It is an open question whether GPGPUs can be exploited to support general irregular parallel workloads [9]. Although prior work has achieved good performance on a limited set of irregular parallel workloads on GPUs [4], such advanced techniques are difficult to program. In this work, we seek to solve the problem of irregular parallelism more generally.

One programmer-friendly approach to irregular parallelism is task parallelism. In the task-parallel programming model, the programmer specifies the generation of new tasks and dependences between tasks. Most task-parallel applications are written to work with a task-parallel runtime that provides a set of operations (*e.g.*, fork task, join, etc.) and manages the parallel tasks. Without a runtime, programming a task-parallel application requires the programmer to be an expert in the intended hardware architectures (e.g., GPU), parallel programming, and algorithm design. With a runtime, the programmer can simply focus on how to express task parallelism while the runtime does everything else—schedules tasks, balances load, and enforces consistency between dependent tasks.

Task-parallel runtimes exist for CPUs, with the most notable example being Cilk [2][18], but runtimes targeting CPUs are a poor fit for GPUs. To understand why this mismatch exists, we must first understand the performance of an idealized task-parallel application (with no runtime) and then how the runtime's overhead affects it. The performance of a task-parallel application is a function of two characteristics: its total amount of *work* to be performed ($T_1$, the time to execute on 1 processor) and its *critical path* ($T_\infty$, the time to execute on an infinite number of processors). Prior work has shown that the runtime of a system with $P$ processors, $T_P$, is bounded by $T_p = O(\frac{T_1}{P}) + O(T_\infty)$ due to the greedy offline scheduler bound [3][10].

A task-parallel runtime introduces overheads and, for purposes of performance analysis, we distinguish between the overheads that add to the work, $V_1$, versus the overheads that add to the critical path, $V_\infty$. When a runtime's overheads are included, the execution time becomes: $T_p = V_1 \frac{T_1}{P} + V_\infty T_\infty$. When designing a runtime, one must be aware of how the design affects these overheads. For example, in a runtime with a global task queue, pulling a task from this queue could require a lock before work can be executed, which increases $V_1$. By contrast, a runtime with a local task queue can avoid the lock, but then the runtime's extra load-balancing work increases $V_\infty$.

Runtime designers must be aware of how their designs affect these overheads. Cilk-5 [8] introduced the "work-first principle" that strategically avoids placing overheads in the work even at the cost of additional overhead on the critical path. Adding to the critical path may seem counter-intuitive, but the equation for $T_p$ above reveals that, in the common case when $\frac{T_1}{P} \gg T_\infty$, the total work dominates the critical path and the work overhead thus has more impact than the critical path overhead.

The work-first principle inspired Cilk-5's work-stealing task scheduler, but a similar implementation is a poor fit for GPUs. A work-stealing task scheduler requires fine-grain communication between threads, which is acceptable on a CPU but expensive on GPUs. In particular, protecting a local task queue from thieves (threads trying to steal work) requires locks and fences. Locks and fences degrade the performance of a GPU's memory system and thus slow down the execution of work. As a result, the use of locks and fences *on a GPU* violates the work-first principle because $V_1$ and $V_\infty$ are coupled. For GPUs, we want to adapt the work-first principle so that implementations can avoid operations that are expensive on GPUs.

*We propose the work-together principle to decouple the runtime overheads on the work and the critical path*. The work-together principle extends the work-first principle to state that (a) the overhead on the critical path should be paid by the entire system at once and (b) work overheads should be paid co-operatively. The work-together principle discourages the adversarial design of a work-stealing task scheduler. Instead, we propose a new work-together task scheduling technique that executes efficiently on a GPU. In particular, the scheduler takes advantage of the GPU's hardware mechanisms for bulk synchronization.

We implement this work-together principle in our *Task Runtime with Explicit Epoch Synchronization (TREES)*. In TREES, computation is divided into massively parallel epochs that are synchronized in bulk. The sequence of epochs is the critical path of a task-parallel program in TREES. TREES provides an efficient and high-performing backend for task-parallel programs that works well on current GPUs. TREES can handle the execution of a vast number of tasks with complex dependency relationships with little runtime overhead. TREES helps to achieve the theoretical speedup, $\frac{T_1}{T_p} = O(P)$, that a *P*-processor GPU could provide for a task-parallel algorithm.

In this work we make the following contributions:
- We propose the work-together principle to guide the design of task-parallel runtimes on GPUs.
- We develop and evaluate a new runtime system, TREES, that efficiently supports task parallelism on GPUs using the work-together principle.
- We have publicly distributed TREES. (For blind review, we do not provide the link nor does the public repository use the "TREES" name.)

## 2. Background

We now present background on fork/join task parallelism and Cilk-5's work-first principle. Our goal is to extend fork/join task parallelism to GPUs by extending the work-first principle in a GPU-aware way.

### 2.1. Fork/Join Task Parallelism

Fork/join task parallelism makes it easy to turn a recursive algorithm into a parallel algorithm. Independent recursive calls can be forked to execute in parallel, and join operations wait for forked tasks to complete before executing. In divide-and-conquer algorithms, forks and joins often correspond to divides and conquers, respectively.

Any task-parallel runtime incurs overhead for the implementation of a fork operation in both the work and critical path. Streamlining fork operations is important because all work that is performed must first be created.

Because a join must be scheduled after all forked operations complete, scheduling a join will incur overhead to ensure the completion and memory consistency of forked tasks. This consistency and notification of completion has the potential to add overhead to the work.

### 2.2. Work-First Principle

Cilk-5's [8] work-first principle states that a runtime should avoid putting overhead on the work, even if that results in overhead on the critical path. Because the algorithm's parallelism ($\frac{T_1}{T_\infty}$) is generally much greater than the hardware can provide (*P*), applying overhead to the work is extremely costly. As long as $V_1 \frac{T_1}{P} \gg V_\infty T_\infty$, the runtime can tolerate critical path overheads and still provide near-linear speedups (i.e., nearly $\frac{T_1}{P}$). This linear speedup is competitive with a greedy offline schedule: $T_P = O(\frac{T_1}{P}) + O(T_\infty)$. Because the work-first principle applies to an online system, the greedy offline schedule is an optimal lower bound on execution time.

The work-first principle inspires Cilk-5's task scheduler. Each processor has its own task queue; a processor pushes forked tasks to the head of its queue and pulls tasks from the head of its queue. If a processor runs out of work in its own task queue, it steals a task from the tail (not head) of another processor's queue. Work-stealing thus balances the load across the processors.

Cilk-5's particular work-stealing task scheduler is carefully implemented to adhere to the work-first principle. Because pushes and pulls are from the head of the queue and steals are from the tail, Cilk-5 incurs synchronization overheads for fork and join (which would constitute undesirable work overheads) only when a thief actually contends with its victim. Synchronization via locking is required only for stealing (and in the relatively rare case of a pull from a queue of size one, *i.e.*, the head and tail are the same).

By design, Cilk-5's runtime overhead is incurred by thieves attempting to steal work but not by the tasks themselves. The number of steal operations is bounded by the product of the number of processors and the critical path ($O(PT_\infty)$). Furthermore, because this work-stealing overhead is performed in parallel, the overhead for stealing is bounded by the critical path ($O(T_\infty)$).

Cilk-5's work-stealing task scheduler is an efficient design for CPUs. However, it requires fine-grain communication and synchronization between threads, including locks and fences, and GPUs are notoriously inefficient at performing these activities.

## 3. Work-Together Principle

We propose the work-together principle which, like the work-first principle, is intended for runtime systems that support fork/join parallelism. However, unlike the work-first principle—which is primarily intended for CPUs—the work-together principle reflects a GPU's strengths and weaknesses.

<u>Strengths:</u> A GPU performs well when it can keep its vast number of hardware thread contexts highly utilized, and this happens when the GPU performs the same operation across a vast number of threads. If the operations access memory, the GPU performs well if the accesses can be coalesced.

<u>Weaknesses:</u> A GPU performs poorly when its hardware thread contexts are under-utilized. Even if there is plenty of work available, under-utilization can occur when the actions of one thread interfere with the performance of other threads. A GPU's SIMT execution model couples the performance of each thread to other threads, and this coupling is problematic when either thread behavior diverges (*e.g.*, threads follow different branch paths or threads access memory in patterns that cannot be coalesced) or when threads must synchronize (*e.g.*, with atomic read-modify-write operations or fences[1]). Thus, even if a work-first runtime adds overheads to a thread's critical path, that overhead can interfere with the work of other threads.

The work-together principle extends the work-first principle—put overhead on the critical path when possible—with two tenets:

- Tenet 1: Pay critical path overheads at one time, in bulk, to minimize interfering with the execution of work.
- Tenet 2: If work overhead is inevitable, cooperatively incur this overhead to reduce the impact on execution time.

By following the work-together principle, a task-parallel runtime can achieve good performance on a GPU, because the hardware provides efficient mechanisms for bulk-synchronous operations (Tenet 1) and SIMT computations and coalesced memory operations (Tenet 2). The work-together runtime's bound is the same as for work-first: $T_P = (V_1 \frac{T_1}{P}) + V_\infty T_\infty$.

### 3.1. Satisfying Work-Together Tenet 1

Extending the work-first principle with Tenet 1 is crucial for performance. Putting overheads on the critical path, rather than the work, is still preferable, but we must incur these overheads in bulk rather than one at a time. Incurring critical path overheads in bulk allows hardware with high thread-level parallelism (*e.g.*, a GPU) to amortize the overheads across multiple threads. Ideally, the runtime would perform forks and joins in bulk rather than at arbitrary times. In this case, the runtime would adhere to Tenet 1, and its overhead on the critical path would neither vary with the number of cores nor interfere with the work.

Although the runtime wants to incur all critical path overheads at once, the programmer expects to be able to fork and join at arbitrary times. The typical task-parallel programming interface allows any thread to fork or join whenever it is ready to do so.

To resolve the tension between what the runtime wants (bulk fork/join) and what the programmer wants (arbitrary fork/join), we propose a runtime that provides the interface the programmer wants while, "under the hood", the runtime satisfies Tenet 1. As a model of such a runtime, we have designed an abstract machine (presented in Section 4) called the Task Vector Machine (TVM). The TVM expands parallelism in a breadth-first manner and executes each level of the generated task dependency graph in a bulk-synchronous manner. The CPU schedules tasks onto the GPU in bulk (*i.e.*, launching kernels with hundreds of tasks at a time) in a manner that respects the inter-task dependencies of the fork/join model. These runtime operations are all paid at once along the critical path. The TVM's bulk synchronous design enables the GPU's hardware to balance the load of tasks across the cores, rather than have to rely on software for this purpose. Furthermore, without fine-grain synchronization, the TVM avoids work overheads due to interference caused by memory fences and atomic operations.

### 3.2. Satisfying Work-Together Tenet 2

Any runtime that implements the TVM will satisfy Tenet 1, but the runtime's implementation details determine whether the runtime also satisfies Tenet 2. The runtime implementation must be tuned to a GPU's strengths, and we want the runtime to shield the application programmer from having to reason about the GPU hardware.

Our runtime, TREES (presented in Section 5), ensures that its work overhead is performed efficiently on GPU hardware. As much as possible, TREES uses SIMT computations and coalesced memory accesses. TREES also tries to minimize the use of atomic operations.

## 4. Task Vector Machine (TVM): A Work-Together Abstract Machine

The work-together principle provides the theoretical ideal for a high-performance task-parallel runtime for a GPU. However, to be useful, the work-together principle must be realized in an actual runtime. Before we present an implementation of the runtime, we first

---

[1] Fences are particularly bad because implementations often flush caches and halt execution for work-items that share a core.

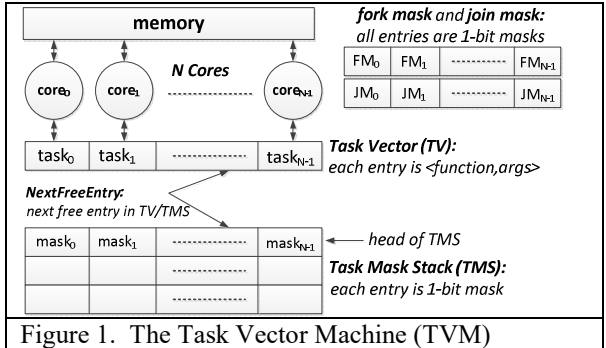

Figure 1. The Task Vector Machine (TVM)

present an abstract machine, the Task Vector Machine (TVM), that obeys the work-together principle.

We introduce this abstraction first for two reasons. First, it provides a means of understanding the key mechanisms of a work-together runtime without the implementation details. Second, this abstract machine provides a mathematical formalism of the model of computation. We use this formalism as a basis for reasoning about the time and space overheads inherent in the model of computation. Ultimately, the TVM serves as the basis for the implementation of our work-together runtime, TREES, presented in Section 5.

The TVM, which we illustrate in Figure 1, has $N$ abstract cores and contains an $N$-wide *Task Vector* (TV). Each core has its own entry in the TV, which is a task that is described with a <function name, arguments> tuple. The TVM has a stack of $N$-wide (one per core) execution bit-masks, called the *Task Mask Stack* (TMS). The TVM also has three bookkeeping structures whose uses are described in more detail later: *NextFreeEntry* points to the next free-to-be-allocated entry in the TV/TMS, and the *fork mask* and *join mask* track requested forks and joins.

The TVM executes an application as a series of *epochs*. In each epoch, every core executes its task in the TV, predicated on its bit at the head of the TMS. During the execution of each epoch, a task mask is popped and task masks may be pushed.

### 4.1. TVM Interface and Programs

The TVM's computational model naturally aligns with task-parallel style programming languages, such as Cilk [2]. One could implement a Cilk compiler to target a TVM-based execution model (such as TREES, the TVM-style runtime for GPUs which we propose in Section 5). To avoid the implementation complexities of writing or modifying a full Cilk compiler, we use a simpler Cilk-like language that supports the key features and primitives of the TVM. Programs written for this language require explicit continuation passing [19] like the original Cilk. In the future, we could provide support for compiling Cilk, or a variety of other task-parallel languages, to work with TREES.

Figure 2 (shown later, where it is discussed more) gives an example program in our Cilk-like language.

The code can be used to traverse a tree in preorder or postorder.

### 4.2. Data-Parallel Extension: Map

This basic TVM interface is sufficient but, because it does not leverage the GPU's SIMD hardware, it does not fully adhere to the work-together principle. We extend this basic TVM interface to include a data-parallel **map** operation that leverages the GPU's high-bandwidth and low-latency memory for groups of threads (*e.g.*, a work-group in AMD/OpenCL platforms). The **map** operation launches a data-parallel task that, compared to a basic **fork** operation, incurs a much smaller overhead per amount of computation.

### 4.3. TVM Execution Model

The initial state of the TVM reflects the number of (abstract) cores in the machine, as well as the initial task (e.g., <postorder, root> for the first call in a postorder traversal of a tree). For simplicity, we assume the number of cores, $N$, is specified as part of the TVM and is fixed for its execution.

The initial task is assigned to core 0 by placing the task's information in entry 0 of the TV. The TMS is initialized to have "true" in entry 0, indicating that the task in entry 0 of the TV should now run on core 0. The TMS is initialized to "false" in all other entries, indicating that no other core should run any task now. The NextFreeEntry is then set to 1, indicating that the next task to be forked should be placed in TV entry 1.

Once the TVM is initialized, execution proceeds as a series of epochs. Each epoch is divided into three phases that are serialized with respect to each other (and with respect to other epochs); however, the computations within Phase 2 are performed in parallel. Phases 1 and 3 are quite short, so there is no need to parallelize them.

#### 4.3.1  Epoch Phase 1 (Epoch Setup)

The first phase of an epoch sets up three bit-vectors that comprise the state required to control the rest of the epoch. The first bit-vector is the task mask. This bit-vector, which controls which tasks execute during the epoch, is initialized by popping the top of the TMS. (Recall that, in the first epoch, the task mask will have a 1 for core 0, and a 0 for all other cores.) As the program forks and joins tasks later in the epoch, the TMS is manipulated such that the TVM executes the proper tasks—those that have been forked and are not waiting on joins of unfinished tasks—in any given epoch. The other two bit-vectors, the fork mask and join mask, support this manipulation. They are initialized to all-zero now, and set by forks and joins, as we see shortly.

### 4.3.2 Epoch Phase 2 (Execute Tasks)

The active cores (*i.e.*, cores with a 1 in the corresponding entries in the task mask) run the tasks assigned to them (*i.e.*, the tasks in the corresponding entries of the task vector). The tasks comprise a mixture of "simple computation"—the normal computational operations, such as arithmetic and memory accesses—and the TVM's primitives for task parallelism: **fork**, **join**, **emit,** and **map**.

The **fork** operation (**fork** f(arg)) spawns a new task to execute, f(arg). However, to obey the work-together principle, the new task is not eligible for immediate execution. Instead, the earliest that it can execute is the start of the next epoch. Spawning a task requires manipulation of the task vector and the fork mask. Although these updates may seem to be inherently serializing (because they access a shared resource), no locks are required. Instead, an atomic increment of NextFreeEntry allocates an entry in the TV and specifies which entry in the fork mask to manipulate. The core performing the **fork** then writes the forked task into the appropriate entry in the TV with no locks or fear of races. The core also sets the corresponding bit in the **fork mask**, indicating that the task in that position is newly forked during this epoch.

The **join** operation (**join** f(arg)) schedules f(arg) to execute *after* the completion of any tasks forked by the current task. Scheduling this **join** requires that the core perform three operations. First, it must replace its corresponding TV entry with <f,arg>, since that is what this core must execute in a later epoch. Next it must schedule that task to run after all forked tasks complete. While this scheduling may seem hard to accomplish in a manner consistent with the work-together principle, the key is that it can be done by the correct manipulation of the TMS at the end of the current epoch. The core sets its corresponding bit in the join mask to request this scheduling and then terminates execution for the current epoch.

The **emit** operation (**emit** value) returns a value from the current task to a parent task waiting to join it. The return result is stored in the task's TV entry, which will not be used further because the task is now done executing. (Correct manipulation of the TMS ensures that the corresponding bit in the task mask is 0 in all future epochs).

The **map** operation (**map** f(arg)) launches a data-parallel task that is executed asynchronously before the next epoch begins.

### 4.3.3 Epoch Phase 3 (Update TMS)

After all tasks have completed execution in a given epoch (i.e., they have either performed **emit** or **join**), the TMS is updated to reflect the forks and/or joins requested during that epoch. Specifically, all forked operations should be scheduled to occur first, followed by all joined operations after the forked operations have finished. Because the TMS is a stack (and thus LIFO), the join mask is pushed first if it has any non-zero bits. (If it is all-zero, no tasks performed joins). Next, the fork mask is pushed if it has any non-zero bits.

If the fork mask is pushed, then the forked tasks will execute at the start of the next epoch, because they are at the top of the TMS. Any forks or joins performed by these children tasks will be pushed onto the TMS on top of the join mask (if any), ensuring that the forks and joins of the children tasks will happen before the joins of the current tasks and thus ensuring that the desired semantics of the **join** operation are maintained.

It is possible that both the fork mask and join mask are all-zero at the end of the epoch (*i.e.*, for "leaf tasks" or **join** tasks that do not spawn any children). If the TMS becomes empty, then the TVM halts as it has performed all requested computations. If the TMS is not empty at the conclusion of the third phase, then the next epoch begins and the cycle repeats.

### 4.4. TVM Analysis

The TVM's execution model highlights how it adheres to Tenet 1 of the work-together principle. Task scheduling is performed in bulk on an epoch-by-epoch basis. Tasks created by forks and joins do not execute as soon as they are ready but instead execute in batches in an epoch.

#### 4.4.1 Time Complexity

The TMS's execution model reveals the critical path and work of a task-parallel application. Each execution mask in the TMS corresponds to an epoch, and the sequence of epochs constitutes the application's critical path. The number of valid bits in the TMS, summed over the entire execution, is the work (as measured in tasks executed). Because the TVM is an idealized machine with $O(T_1)$ cores, the execution time is $O(T_\infty)$.

In practice, the TVM is designed to be implemented as a runtime for GPUs. We approximate the performance of a GPU as a TVM with $P$ processors that are each $W$-wide SIMD. If we pessimistically assume a 50/50 branch split and an average thread divergence penalty of $log(W)$, then this system would have an execution time of $T_{P,W} = V_1 \frac{\log(W)T_1}{PW} + V_\infty T_\infty$. The best-case execution time is when the SIMD width executes without divergence and $T_{P,W} = V_1 \frac{T_1}{PW} + V_\infty T_\infty$. In the worst case, the execution time can be upper bounded by the maximum nesting of branches ($D$) where ($2^D < W$), in which case $T_{P,W} = V_1 \frac{2^D T_1}{PW} + V_\infty T_\infty$.

*4.4.2 Space Complexity*

Each TVM core requires space for a task and a stack of mask bits. As a result, the space complexity of an algorithm running on the TVM is upper-bounded by the work ($O(T_1)$) and lower bounded by the parallelism ($\Omega(\frac{T_1}{T_\infty})$). The upper bound holds because each task can only use one function, one set of arguments, and one stack of bits that each requires memory to compute the work. The lower bound holds because there needs to be at least one function, one set of arguments, and one stack of bits for each active task (parallelism).

## 5. TREES: TVM on GPUs

The TVM provides an appealing model of computation for task-parallelism on a GPU-based system, but to be useful, it must be implemented in a real runtime. We design such a runtime, called Task Runtime with Explicit Epoch Synchronization (TREES), which targets CPU/GPU hybrid systems. A hybrid system allows TREES to use the CPU for the serial portions of the computation (*e.g.*, epoch setup), and execute the parallel computations on the GPU.

A programmer writes TREES source code, which resembles C++ with the task-parallel primitives provided by the TVM execution model: **fork**, **join**, **emit**, and **map**. The programmer then uses our TREES compiler, which translates the code into OpenCL and includes the TREES runtime. We chose OpenCL as the backend heterogeneous language because of its portability, but other options exist.

One may expect that TREES is simply a direct implementation of the TVM in OpenCL; however, this naïve implementation would perform poorly. The TVM has many facets that are convenient for an abstract formalism, but ill-suited to a real implementation, such as an arbitrary number of abstract cores. Furthermore, TREES can use the highly-parallel nature of GPUs to amortize costs across multiple work-items (threads), adhering to Tenet 2 of the work-together principle.

### 5.1. TREES Structures

Although TREES *logically* contains the same structures as the TVM, its *physical* realization of those structures often differs. In particular, we design TREES to facilitate SIMT execution and coalesced memory accesses.

*5.1.1 Cores*

Although the TVM model allows an arbitrary number of *abstract* cores, a real implementation is inherently constrained by the actual number of cores in the GPU hardware (*e.g.*, 256). Even though only some elements of the Task Vector execute in any particular epoch—specifically, those whose corresponding Task Mask entries are "true"—there may be many more tasks ready to execute than the GPU has physical cores.

Fortunately, the GPU's hardware scheduler is designed to handle this situation. TREES represents each TVM core as an OpenCL work-item and each epoch as a kernel launch with an NDRange (set of work-items) that includes the tasks that can execute in the current epoch. TREES then splits this NDRange into work-groups of 256 work-items and relies on the GPU's hardware scheduler to schedule the work-items to the GPU's cores.

*5.1.2 Task Vector and Task Mask Stack*

In the TVM, the TV is an array of <function, arguments> tuples, and the TMS is a stack of bit-vectors. However, to enable memory coalescing when reading or writing the TV and TMS, TREES uses different data structures, all of which reside in the GPU's memory space.

The TVM's representation of the Task Mask Stack is inefficient for a real implementation. One inefficiency with this representation is the memory bandwidth required at the end of each epoch to manipulate the TMS. Each TMS entry is a bit-vector with a number of bits equal to the maximum number of possible tasks, and up to two such masks (fork and/or join) may be pushed onto the TMS at the end of each epoch. A second inefficiency is the memory and compute bandwidth required to find the active tasks at the start of each epoch. Not only must the bit-vector corresponding to the TMS be read in from memory, but also each bit must be examined to determine which tasks are active. This computation poses a further difficulty in that it seemingly must be performed by the CPU, as it is required to determine what to launch to the GPU for a particular epoch. Launching all possible tasks to the GPU, then checking if the tasks are valid would work correctly, but performance would suffer due to poor utilization.

TREES uses a different representation motivated by three observations. First, in any column of the TVM's Task Mask Stack, there is at most one "true" bit at any given time. Consequently, TREES encodes the information about when a task should be executed with a single Epoch Number (EN) instead of a stack of bits. A task's Epoch Number corresponds to the height on the stack of the task's "true" bit (counting from one at the bottom of the stack), or 0 if the task has no "true" bits. A task can execute if its EN matches the current epoch number (**CEN**).[2] The CPU explicitly maintains

---

[2] More precisely, each TV entry holds one integer to encode both the task type *and* the epoch number in which it executes. The integer that specifies "a task executes function number *taskType* in epoch number *someEpoch*" is encoded as

the CEN and manages which epoch number needs to be executed next. The CPU uses a kernel argument to pass the **CEN** to the GPU.

The second observation is that tasks with the same Epoch Number are generally contiguous in the TMS (and TV). Recall that tasks are entered into the TV/TMS either by **fork**, which allocates a new entry at the end of the TV/TMS, or by **join**, which replaces the current entry in the TV/TMS. Consequently, TREES can track the range of indices in which tasks with a given epoch number can be found.

The third observation is that explicit bit-vectors are not needed to track forks and joins. Instead, TREES tracks the next free TV/TMS index with a single integer (**nextFreeCore**), which is incremented atomically by a thread performing a **fork**. At the end of the epoch, the CPU can compare the **nextFreeCore** index to its value before the epoch started (**oldNextFreeCore**) to determine if any forks have happened. As with forks, for joins, the important determination to make is whether or not any joins have been scheduled during an epoch. TREES implements this determination with a single bit, **joinScheduled**, which is cleared by the CPU before the epoch starts, set by any thread performing a join, and read by the CPU at the end of the epoch. Similarly, **mapScheduled** is cleared by the CPU before the epoch starts, set by any thread performing a **map**, and read by the CPU at the end of the epoch.

## 5.2. TREES Execution

TREES executes programs according to the semantics of the TVM model. Unlike the abstract TVM, TREES must consider the performance implications of specific implementation decisions. Because TREES is targeted at CPU/GPU systems, one important decision is whether the CPU or GPU performs a specific portion of the computation. For computations performed on the GPU, TREES must exploit the GPU's SIMT nature to maximize performance.

### 5.2.1 Initialization

The initialization of an application is largely the CPU's responsibility. The CPU initializes the basic state of the TREES runtime—creating an empty join stack, initializing the CEN to 0, etc. The CPU does not initialize the entire Task Vector, as that resides in GPU memory. The CPU writes the application's initial task in the TV's initial ($0^{th}$) entry, which will run in epoch 0.

---

*someEpoch\*NumTaskTypes+taskType*. An entry in the range between *CEN\*NumTaskTypes+1* and *(CEN+1)\*NumTaskTypes*, inclusive, denotes a task that executes in the current epoch.

### 5.2.2 Epoch Phase 1 (Epoch Setup)

The setup at the start of each epoch is also the CPU's responsibility. Much of the work is serial manipulation of the book-keeping variables: setting **oldNextFreeCore** to the current value of **nextFreeCore**, resetting **mapScheduled** and **joinScheduled** to 0, and popping the join stack and NDRange stack. The CPU then sets up an OpenCL kernel—which is part of the TREES runtime—to run the appropriate tasks from the TV. Once the CPU has enqueued the kernel, epoch setup is complete.

### 5.2.3 Epoch Phase 2 (Execute Tasks)

The task execution phase, which is the highly parallel portion of the computation, is performed by the GPU. The CPU spends phase 2 waiting for the GPU to complete its kernel. Each work-item in the current NDRange (set by the CPU during Epoch Phase 1) starts with TREES runtime code that reads the appropriate TV entry, determines which application function to execute, and calls that function with its argument (which is in the TV entry). The work-items execute their tasks, which contain "normal" computational code and the TVM's task-parallel primitives.

Executing the task-parallel primitives in an efficient fashion is the key to TREES performance. These primitives appear to require operations on which GPUs perform poorly: synchronized accesses to various data structures. Fortunately, the structural design of TREES allows these operations to be performed with atomic increments or set-to-1 operations rather than locks. For example, when a core performs a fork, it obtains a new slot in the TV by atomically incrementing **nextFreeCore** using a local memory reduction to ensure a single atomic operation per wavefront. The core—and the other cores executing the same task type in SIMT fashion, which have also obtained new slots in the TV—perform one coalesced write to the TV to set the task type and additional coalesced writes to set each of the arguments to the function that was forked.

Performing a map operation is similar, except that the core(s) executing it sets **mapScheduled** to 1 after manipulating the TV. Likewise, when a core performs a join operation, it (and other cores executing in a SIMT fashion with it) set **joinScheduled** to 1. As with fork, cores performing join or emit operations execute coalesced writes to update the TV entries (which are their own entries, not a newly allocated one).

### 5.2.4 Phase 3 (Update TMS)

The end-of-epoch update is performed by the CPU, as this update is a serial computation amenable to the CPU. After the GPU completes the kernel, the CPU enqueues a transfer of **nextFreeCore**, **joinScheduled**, and **mapScheduled** from GPU memory space back to

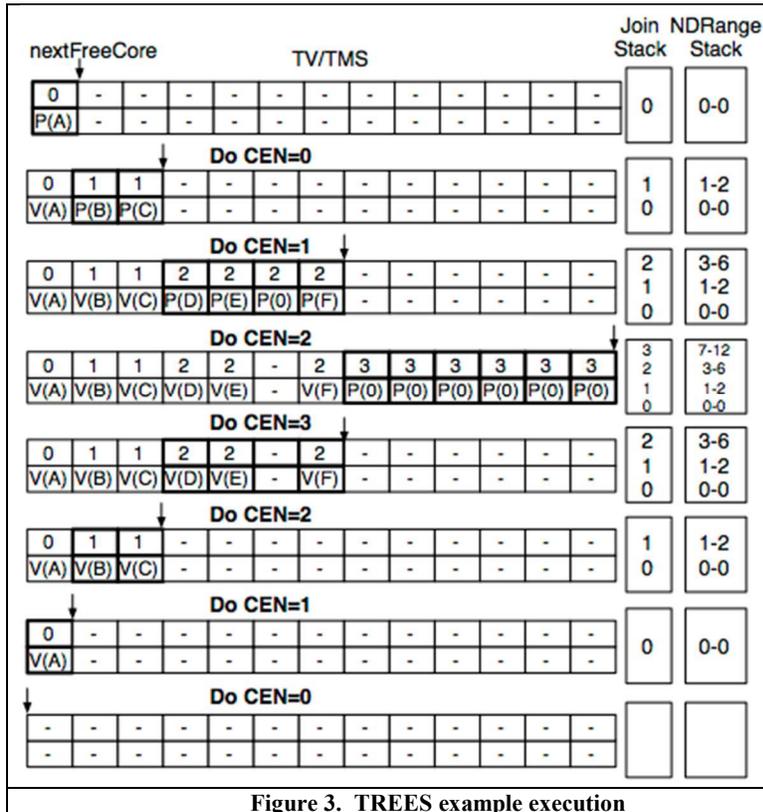

Figure 3. TREES example execution

```
void visit (node) {
    // do visit
}
task void preorder (node) {
    if (node != nil) {
        visit (node);
        fork preorder (node.right);
        fork preorder (node.left);
    }
}
task void postorder (node) {
    if (node != nil) {
        fork postorder (node.right);
        fork postorder (node.left);
        join visitAfter (node) ;
    }
}
task void visitAfter (node) {
    visit (node);
}
```

**Figure 2. Tree traversal on the TVM**

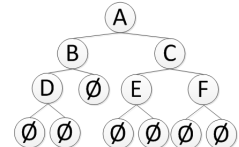

**Figure 4. Example 6-node tree**

CPU memory space. The CPU checks the value of **joinScheduled**; if the value is one, the current epoch number and NDRange are pushed onto the join stack and NDRange stack, respectively. The CPU then compares **oldNextFreeCore** and **nextFreeCore**; if the values differ—indicating that a fork was performed—then CEN+1 is pushed onto the join stack and an NDRange from **oldNextFreeCore** to **nextFreeCore** is pushed onto the NDRange stack. The CPU checks the value of **mapScheduled** and, if it is one, launches a kernel of the data-parallel map operations scheduled during the last epoch. This kernel runs to completion before TREES begins Phase 1 of the next epoch.

*5.2.5  Discussion of Performance Issues*

The hardware scheduler provides load balancing at little cost to the work ($T_1$). Entering the GPU driver to launch the kernel and transferring the shared variables comprise the critical path overhead ($V_\infty$). Trends in GPU hardware and drivers suggest that these overheads will become ever smaller.

**5.3. TREES Example**

Figure 3 shows the step-by-step example of how TREES executes the code in Figure 2 to perform a postorder traversal of the binary tree shown in Figure 4.

In Figure 3, time goes from top to bottom, with the first row depicting the starting state of the TREES runtime, as setup by the CPU. Each row has a top (epoch number) and bottom (task and argument).

After initialization, the only valid entry in the TV/TMS is entry 0, which is scheduled to run in epoch 0, and it has the task of executing the postorder function on argument A (indicated by P(A) in the bottom half of the box). The join stack contains 0, because the only epoch scheduled so far is 0. The NDRange stack contains one entry (0 through 0), representing the TV entry ranges corresponding to this epoch. The active NDRange for the next epoch is indicated by thicker lines around the corresponding entries in the TV/TMS. The **nextFreeCore** is indicated by an arrow, showing the boundary between allocated cores and free cores.

TREES then pops the join stack and NDRange stack and executes an epoch with current epoch number (CEN) of zero. During this epoch, there is only one task to run (postorder on A). This task forks two new tasks, which allocate entries in the TV/TMS; entries 1 and 2 now indicate that they should run postorder of B and C, respectively, in epoch 1. The postorder task on A concludes by scheduling a join (to do visitAfter(A) in the future), replacing the task information in its own entry (which is shown as V(A) in the figure). Because this epoch scheduled a join, its epoch number is pushed onto the join stack, and its NDRange is pushed back onto the NDRange stack. Because this epoch forked new tasks, CEN+1 is pushed onto the join stack and the NDRange spanning the forked TV indices is pushed onto the NDRange stack. The next row of the figure shows the state of the TREES runtime at the conclusion of this epoch.

Epochs 1 and 2 proceed in much the same way; the appropriate tasks run, fork new tasks, and schedule joins. The one notable difference in epoch 2 occurs with the task executing postorder(NULL). This task does not schedule a join, so it does not update its TV entry to indicate future computation. Instead, when it is done, it marks its TV entry as invalid. Even though this entry is invalid, it cannot be re-used yet, as allocation is only performed by manipulating **nextFreeCore**.

Epoch 3 behaves differently from the previous epochs, as no tasks fork new tasks nor schedule joins (they all execute postorder(NULL)). Consequently, all corresponding TV entries (7-12) are marked as invalid, and **nextFreeCore** decreases, allowing these entries to be re-used as needed. The NDRange stack and join stack decrease in size for the first time (nothing is pushed) in our example, revisiting previously used TV entries to execute the tasks they have scheduled after their child tasks have completed. As always, the value popped from the top of the join stack (currently 2) becomes the CEN for the next epoch; the difference now is that this change causes the CEN to decrease.

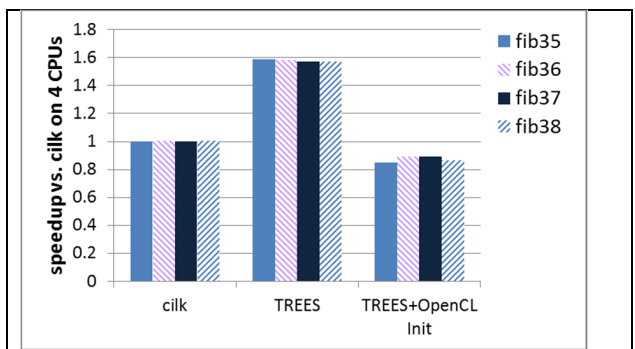

**Figure 5. Performance on Fibonacci**

When TREES now executes with CEN=2, there are 4 tasks in the NDRange, of which only 3 are valid. The CPU will launch all 4, each of which will read their TV/TMS entry to determine what to do. The three valid tasks will all perform the visitAfter function on their respective arguments. The other task will determine it is invalid by reading its TV/TMS entry and exit immediately.

The remaining epochs proceed in a similar fashion—executing visitAfter tasks, popping the join and NDRange stacks, and decreasing **nextFreeCore**—until the join stack, NDRange stack, and TV/TMS become empty (which are all guaranteed to occur at the same time). At this time, the TREES application is complete.

### 5.4. Why TREES is Work-Together

We list the key design aspects of TREES that enable it to satisfy Tenet 2 of the work-together principle.
- Reads/writes to the TV/TMS are coalesced memory accesses.
- Forks and joins are performed in a SIMT fashion.
- There are no fences. Synchronization that would require fences on the GPU is performed by the CPU.
- Atomic RMW instructions are used rarely.

- Cores that perform the same task types tend to create tasks of the same type, and the created tasks tend to run on contiguous cores in a SIMT fashion.

### 6. Experimental Evaluation

We answer four questions with our experiments:

1) Does task-parallelism on GPUs ever make sense? That is, are there task-parallel applications that run faster on a GPU than on a CPU? (Section 6.2)

2) Can the overhead of a task-parallel runtime be small? (Section 6.3)

3) Is the data-parallel **map** operation important to TREES performance? (Section 6.4)

4) Based on experiential evidence, is it reasonably easy to program to the TVM interface? (Section 6.5)

### 6.1. Experimental Methodology

Our hardware platform is an AMD A10-7850k APU that is well-suited to TREES. This APU is the first chip to support shared virtual memory that allows for low-latency kernel launch and memory transfers. The trends in CPU/GPU chip design, exemplified by the HSA standard [20], suggest that future chip designs will be even more suitable for TREES.

We run all experiments with the Catalyst 14.1 drivers on the Linux 3.10 kernel with OpenCL 1.2.

### 6.2. Task Parallelism on GPUs

The first question we explore is whether there is performance potential from running task-parallel applications on GPUs. If CPUs always outperform GPUs in a task-parallel setting, then designing runtimes for GPU task parallelism is not terribly useful.

We study two extremes of task-parallel applications. The first application uses a naive algorithm for calculating Fibonacci numbers, in which the application performs virtually no computation per task. This application is a worst-case scenario for TREES in that it maximizes the ratio of runtime overhead to useful work. The second application calculates an FFT, and its tasks perform a significant amount of computation. This application is a much better scenario for TREES in that the ratio of runtime overhead to useful work is low.

Figure 5 shows the results for Fibonacci as a function of which Fibonacci number is computed (35-38). We plot the speedup of TREES—with and without OpenCL initialization latency—with respect to Cilk on four CPU cores. We separate out the one-time OpenCL initialization latency for the application because, in an application like Fibonacci that performs minimal computation, OpenCL initialization overheads are relatively large. Also, trends in CPU/GPU chip design suggest that OpenCL overheads will decrease in the future. We see that, if we ignore OpenCL overheads, TREES outperforms the parallel performance of Cilk. Because relative performance does not vary with problem size, TREES balances load similarly to the Cilk-5 runtime. When considering OpenCL overheads, TREES performs somewhat worse than Cilk.

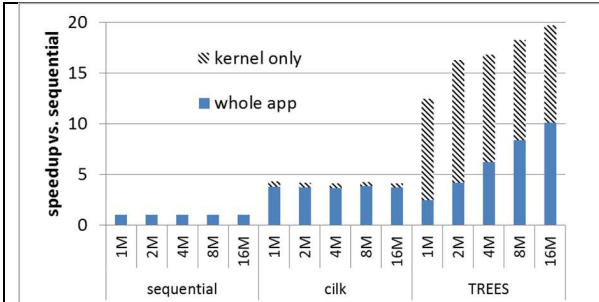

Figure 6. Performance of FFT

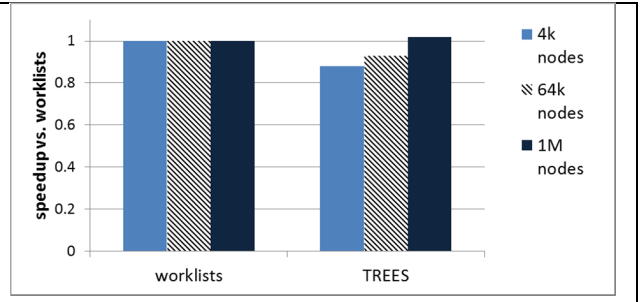

Figure 7. Performance of BFS

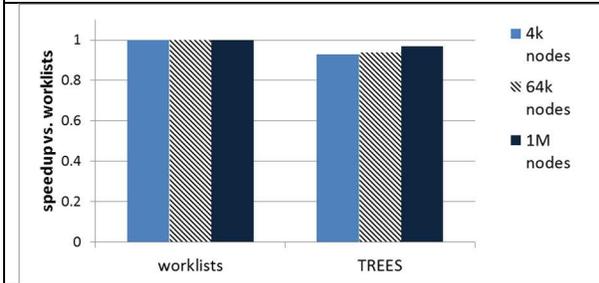

Figure 8. Performance of SSSP

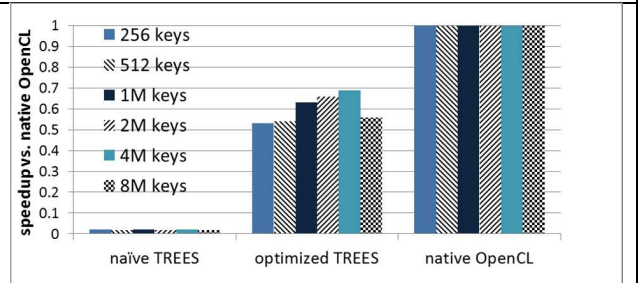

Figure 9. Performance of sort

For FFT, Figure 6 shows the speedups, with respect to the sequential implementation, for both the whole program and just the parallel kernel. We do not use data-parallel **map** operations, which would benefit TREES. When excluding initialization costs, TREES always outperforms the sequential and Cilk versions. When including initialization costs, an FFT must be larger than 1M to see a benefit from using the GPU.

The experiments show there is potential for at least some task-parallel applications on GPUs. Furthermore, we expect that TREES will do even better on future chips with even tighter integration between the CPU and GPU that will reduce kernel launch latencies.

### 6.3. TREES Runtime Overhead

TREES provides programmability for general task-parallel applications. One might suspect that its generality incurs some performance cost. To study this overhead for generality, we compare applications on TREES to native OpenCL versions of the same applications.

An interesting set of applications are graph algorithms, because there is both recent work in this area and existing benchmarks. The Lonestar benchmark suite [13] and its related LonestarGPU suite use work-lists to implement bfs (breadth-first search) and sssp (single-source shortest-path). These benchmarks use input and output work-lists to allow efficient push and pull operations. The pull operation is data-parallel on the input worklist. Pushing to the output work-list uses a single tail pointer that is atomically incremented with new vertices to explore. After a kernel execution has completed, the host transfers a single int to see if a new relaxation kernel is necessary. If so, the input and output work-lists are swapped and the launch bounds of the next kernel are determined by the size of the old output worklist. This execution continues until no new nodes are explored during a relaxation kernel.

In many ways, the techniques used in these Lonestar benchmarks are a hand-coded subset of the TREES implementation. Fundamentally, the two perform the same basic algorithm. Whereas the Lonestar programmer manages the work-lists by hand, the TREES runtime manages it in the Task Vector.

We ported the LonestarGPU implementations of bfs and sssp from CUDA to OpenCL, and we compared them to TREES. Figure 7 and Figure 8 show the results for bfs and sssp, respectively. We can see that TREES is never more than 6% slower than the LonestarGPU equivalent benchmark. The performance difference likely comes from the extra effort to determine the task type of the bfs and sssp.

The results show we are paying minimal cost for the generality of TREES. Because we studied only the runtime on the GPU (and not the CPU), we are showing the worst-case overhead for TREES. (With the CPU time included, which would be the same for both TREES and native OpenCL applications, the overhead percentage would be less.)

### 6.4. Data-Parallel Map

In this section, we study the use of data-parallel **map** operations to optimize a task-parallel sort. This application represents an atypical use of TREES, because the available parallelism is highly regular. However, this application serves to show that data-parallel **map** operations can greatly improve the performance of TREES on task-parallel applications that also contain regular data parallelism.

We compare a naïve TREES implementation of mergesort (without **map** operations), a more sophisticated TREES implementation with **map** operations, and a high-performance native OpenCL bitonic sort.

From the results in Figure 9, we first observe that the naïve TREES mergesort performs abysmally. This result is unsurprising, in that this implementation does not exploit the easily available data parallelism. Second, the sophisticated TREES mergesort makes up much of the gap between the naïve TREES mergesort and the native OpenCL sort. This result is perhaps surprising, because the sophisticated TREES mergesort has twice as many kernel launches, more memory copies, and reads arguments from global memory instead of parameter space.

The sophisticated TREES mergesort is still performing only about half as well as the native OpenCL sort, and we can infer that a worst-case performance loss between a native data-parallel program and TREES would be approximately 2-3x.

We conclude that data-parallel **map** operations can be used to exploit the data-parallel GPU hardware and greatly improve performance. On applications that are better-suited to TREES—sort is an extremely bad match for TREES that we use for illustrative purposes—the **map** operations could make the difference between good and poor performance.

### 6.5. Programmability

We experientially analyzed the programmability of TREES with undergraduate programmers with no parallel programming experience. Many of the undergraduates had taken only undergraduate computer architecture and an introduction to data structures and algorithms. Our "results" here are obviously not scientifically rigorous, but they provide a sense of the programmability of the TVM interface.

We found that undergraduates were easily able to write the following task-parallel applications: nqueens, matrix multiply, traveling salesman, breadth-first search, and simulated annealing. Many of these programs were done with 10-20 hours of work, and most of that work was spent in understanding the algorithms themselves. We have written applications of similar and increased complexity in less time.

Although we do not claim that all applications are easy to write, our experience suggests TREES is a promising platform for general GPU development by the "average programmer."

### 7. Related Work

Inspiring all prior work on task parallelism for GPUs is the large body of work on task parallelism for CPUs, such as the notable examples Cilk [18] and X10 [5].

Orr et al. [17] recently developed an implementation of a task-parallel programming model called channels [9]. Like our work, it enables high-performance task-parallel programming for GPUs. Unlike our work, it requires modifying the GPU's hardware scheduler to help manage the channels.

Some prior work has provided task parallelism on GPUs using "dynamic parallelism." Both CUDA 5.0 and OpenCL 2.0 support dynamic parallelism, in which a GPU kernel can directly launch a new kernel [16][21]. Dynamic parallelism enables task-parallel programming, but with three drawbacks. First, if one writes a single-threaded task, which is how CPU programmers write task-parallel software, the kernel has a single thread and performs poorly. Second, if a parent task is waiting for a child task to complete and the parent task suffers branch divergence, then deadlock can occur [12]. Third, it requires a hardware modification, and most current GPUs do not support it.

Another avenue of prior work in providing task parallelism on GPUs is based on persistent threads [11]. Although the original paper [11] did not propose using persistent threads to support task-parallel programming, subsequent work has done this [6][22]. Like dynamic parallelism, using persistent threads for task parallelism achieves poor performance on the single-threaded tasks used in CPU programming models. Performance also suffers when persistent threads are idle yet contending for resources (e.g., the global memory holding the task queue). Debugging is very difficult, because GPUs require a kernel to complete before providing any access to debugging information. Furthermore, not all hardware supports interrupting execution, in which case a buggy persistent threads program requires a reboot.

The StarPU runtime [1] supports task parallelism on heterogeneous systems, including CPU/GPU systems, but it has a somewhat different goal than TREES or the previously discussed related work. StarPU seeks to provide a uniform and portable high-level runtime that schedules tasks (with a focus on numerical kernels) on the most appropriate hardware resources. StarPU offers a much higher level interface to a system, with the corresponding advantages and disadvantages. The OmpSs programming model [7] extends OpenMP to support heterogeneous tasks executing on heterogeneous hardware. OmpSs, like most prior work, would only achieve good performance on GPUs if the tasks themselves are data-parallel.

Other work has explored the viability of running non-data-parallel, irregular algorithms on GPUs [4][14]. This work has shown that GPUs can potentially achieve good performance on irregular workloads. Interestingly, Nasre et al. [15] studied irregular parallel algorithms from a data-driven rather than topology-driven approach, and this approach uses a task queue to manage work. TREES could complement this work by providing portable support for this task queue.

## 8. Conclusion

We have developed TREES, a task-parallel runtime for CPU/GPU platforms. TREES is based on the work-together principle, which enables it to achieve high performance when using GPUs, and we expect even greater performance from future platforms with tighter coupling between CPU and GPU.

## Acknowledgments

This material is based on work supported by the National Science Foundation under grant CCF-1216695.

## 9. References

[1]     C. Augonnet, S. Thibault, R. Namyst, and P.-A. Wacrenier, "StarPU: A Unified Platform for Task Scheduling on Heterogeneous Multicore Architectures," *Concurr. Comput. : Pract. Exper.*, vol. 23, no. 2, pp. 187–198, Feb. 2011.

[2]     R. D. Blumofe, C. F. Joerg, B. C. Kuszmaul, C. E. Leiserson, K. H. Randall, and Y. Zhou, "Cilk: An Efficient Multithreaded Runtime System," *SIGPLAN Not.*, vol. 30, no. 8, pp. 207–216, Aug. 1995.

[3]     R. P. Brent, "The Parallel Evaluation of General Arithmetic Expressions," *J. ACM*, vol. 21, no. 2, pp. 201–206, Apr. 1974.

[4]     M. Burtscher, R. Nasre, and K. Pingali, "A quantitative study of Irregular Programs on GPUs," in *2012 IEEE International Symposium on Workload Characterization (IISWC)*, 2012, pp. 141–151.

[5]     P. Charles, C. Grothoff, V. Saraswat, C. Donawa, A. Kielstra, K. Ebcioglu, C. von Praun, and V. Sarkar, "X10: An Object-oriented Approach to Non-uniform Cluster Computing," in *Proceedings of the 20th Annual ACM SIGPLAN Conference on Object-oriented Programming, Systems, Languages, and Applications*, 2005, pp. 519–538.

[6]     S. Chatterjee, M. Grossman, A. Sbîrlea, and V. Sarkar, "Dynamic Task Parallelism with a GPU Work-Stealing Runtime System," in *Languages and Compilers for Parallel Computing*, S. Rajopadhye and M. M. Strout, Eds. Springer Berlin Heidelberg, 2013, pp. 203–217.

[7]     A. Duran, E. Ayguadé, R. Badia, J. Labarta, L. Martinell, X. Martorell, and J. Planas, "OmpSs: A Proposal for Programming Heterogeneous Multi-core Architectures," *Parallel Processing Letters*, vol. 21, no. 2, pp. 173–193, 2011.

[8]     M. Frigo, C. E. Leiserson, and K. H. Randall, "The Implementation of the Cilk-5 Multithreaded Language," in *Proceedings of the ACM SIGPLAN 1998 Conference on Programming Language Design and Implementation*, 1998, pp. 212–223.

[9]     B. R. Gaster and L. Howes, "Can GPGPU Programming Be Liberated from the Data-Parallel Bottleneck?," *Computer*, vol. 45, no. 8, pp. 42–52, Aug. 2012.

[10]    R. L. Graham, "Bounds on Multiprocessing Timing Anomalies," *SIAM Journal on Applied Mathematics*, vol. 17, no. 2, pp. 416–429, 1969.

[11]    K. Gupta, J. A. Stuart, and J. D. Owens, "A Study of Persistent Threads Style GPU Programming for GPGPU Workloads," in *Innovative Parallel Computing (InPar), 2012*, 2012, pp. 1–14.

[12]    M. Harris, "Many-core GPU Computing with NVIDIA CUDA," in *Proceedings of the 22Nd Annual International Conference on Supercomputing*, 2008, pp. 1–1.

[13]    M. Kulkarni, M. Burtscher, C. Cascaval, and K. Pingali, "Lonestar: A Suite of Parallel Irregular Programs," in *IEEE International Symposium on Performance Analysis of Systems and Software*, 2009, pp. 65 –76.

[14]    R. Nasre, M. Burtscher, and K. Pingali, "Morph Algorithms on GPUs," in *Proceedings of the 18th ACM SIGPLAN Symposium on Principles and Practice of Parallel Programming*, 2013, pp. 147–156.

[15]    R. Nasre, M. Burtscher, and K. Pingali, "Data-Driven Versus Topology-driven Irregular Computations on GPUs," in *2013 IEEE 27th International Symposium on Parallel Distributed Processing (IPDPS)*, 2013, pp. 463–474.

[16]    NVIDIA, "NVIDIA's Next Generation CUDA Compute Architecture: Kepler GK110." http://www.nvidia.com/content/PDF/kepler/NVIDIA-Kepler-GK110-Architecture-Whitepaper.pdf.

[17]    M. S. Orr, B. M. Beckmann, S. K. Reinhardt, and D. A. Wood, "Fine-grain Task Aggregation and Coordination on GPUs," in *Proceedings of the 41st International Symposium on Computer Architecture*, 2014.

[18]    K. H. Randall, "Cilk: Efficient Multithreaded Computing," Massachusetts Institute of Technology, 1998.

[19]    G. J. Sussman and G. L. Steele Jr., "Scheme: An Interpreter for Extended Lambda Calculus," in *MEMO 349, MIT AI LAB*, 1975.

[20]    V. Tipparaju and L. Howes, "HSA for the Common Man." http://devgurus.amd.com/servlet/JiveServlet/download/1282191-1737/HC-4741_FINAL.pptx, 2012.

[21]    R. Tsuchiyama, T. Nakamura, T. Iizuka, A. Asahara, S. Miki, S. Tagawa, and S. Tagawa, *The OpenCL Programming Book*, 1 edition. Fixstars Corporation, 2010.

[22]    S. Tzeng, B. Lloyd, and J. D. Owens, "A GPU Task-Parallel Model with Dependency Resolution," *Computer*, vol. 45, no. 8, pp. 34–41, 2012.